\documentclass[pra,twocolumn,showpacs,preprintnumbers,%
               amsmath,amssymb,superscriptaddress]{revtex4-1}
\usepackage{amsmath}
\usepackage{graphicx}
\usepackage{dcolumn}
\usepackage{bm}
\usepackage{hyperref}

\begin{document}
\title{Theory of robust subrecoil cooling by STIRAP}
\author{Vladimir S. Ivanov}\email{ivvl82@gmail.com}
\affiliation{Turku Centre for Quantum
Physics, Department of Physics and Astronomy, University of Turku,
20014 Turku, Finland} \affiliation{Saint Petersburg State
University of Information Technologies, Mechanics and Optics,
197101 St. Petersburg, Russia}
\author{Yuri V. Rozhdestvensky}
\email{rozd-yu@mail.ru} \affiliation{Saint Petersburg State
University of Information Technologies, Mechanics and Optics,
197101 St. Petersburg, Russia}
\author{Kalle-Antti Suominen}\email{Kalle-Antti.Suominen@utu.fi}
\affiliation{Turku Centre for Quantum Physics, Department of
Physics and Astronomy, University of Turku, 20014 Turku, Finland}
\date{\today}

\begin{abstract}
  We demonstrate one-dimensional robust Raman cooling in a three-level
  $\Lambda$-type atom, where the velocity-selective transfer is
  carried out by a \mbox{STIRAP} pulse. In contrast to the standard
  Raman method, the excitation profile is insensitive to variations in
  the pulse duration, whereas its position and width are nevertheless
  under control. Two different cooling variants are examined and we
  show that subrecoil temperatures are attainable in both cases.
\end{abstract}


\maketitle

\section{Introduction}
The physics of cold atoms is an important ingredient in the emerging
field of quantum technology, either providing elements for demanding
applications or a transparent testing ground for ideas and
methods~\cite{Mabuchi2002,Bloch2008a,Bloch2008b}. This is made
possible by sophisticated control of both translational and internal
degrees of freedom for neutral atoms with electromagnetic fields,
demonstrated especially by laser cooling and trapping
methods~\cite{Metcalf1999,Weidemuller2009,Cohen-Tannoudji2011}. One of
the most notable successes has been the reaching of quantum
degeneracy, i.e., Bose-Einstein condensation with bosonic atoms, and
Cooper pairing with fermionic
atoms~\cite{Weidemuller2009,Cohen-Tannoudji2011,Pethick2008}. Here the
key tools have been evaporative cooling and magnetic trapping, as
reaching the ultralow temperatures and high densities with laser light
is usually hampered by heating and loss of atoms due to light-assisted
atomic collisions~\cite{Holland1994,Suominen1996} or reabsorption of
scattered photons~\cite{Steane1992,Cooper1994}.

Evaporative cooling, however, is rather wasteful of atoms and requires
fast thermalizing collisions, which limits its practicality in some
cases, such as single-species fermionic
gases~\cite{Duarte2011,MacKay2011}, atomic
clocks~\cite{Ye2008,Derevianko2011}, or small samples of cold
atoms~\cite{Phoonthong2010}, and especially transversal cooling of
atomic beams~\cite{Theuer1999,Vewinger2003}. Very recently, for
instance, fermionic atoms have been efficiently laser cooled using
narrow optical transitions, which lead to low Doppler
temperatures~\cite{Duarte2011,MacKay2011}. The same approach has been
used earlier with alkaline-earth atoms, for the purpose of building
atomic clocks for optical wavelengths~\cite{Ye2008,Derevianko2011}. In
general, state-insensitive traps are needed for quantum state
engineering and precision
metrology~\cite{Ye2008,Phoonthong2010,Derevianko2011}. Similarly, one
needs to develop alternatives also for evaporative cooling. A simple
possibility is to develop further the concept of Raman cooling; one of
the aspects that can be improved is the robustness in respect to the
laser pulse parameters. One of the success stories in robust quantum
control of internal state dynamics is the use of adiabatic following
of eigenstates, i.e., stimulated rapid adiabatic passage
(STIRAP)~\cite{Vitanov2001}. As known e.g. with molecular
dynamics~\cite{Garraway1998,Rodriguez2000}, it can also be used to
control motional degrees of freedom.

Raman cooling~\cite{Kasevich1992} is one the most efficient
non-evaporative techniques for optical cooling of atoms below the
one-photon recoil limit. The cooling efficiency is associated with
population transfer which is typically generated by a $\pi$-pulse that
has a Blackman~\cite{Kasevich1992,Davidson1994,Reichel1994} or a
square~\cite{Reichel1995,Boyer2004} envelope. Although such a
$\pi$-pulse operates efficiently on atoms corresponding to the
resonant velocity group, the process requires extreme control of the
pulse amplitude and duration. Even though some optimization can be
obtained for the method~\cite{Ivanov2011}, an approach based on
adiabatic passage is of definite interest due to its robustness
regarding pulse amplitudes and durations. However, the use of
adiabatic passage in Raman cooling is not possible without having an
exact understanding of its velocity-selective properties. In this
paper we present a careful analysis of the Raman cooling via
\mbox{STIRAP} pulses, and show that subrecoil temperatures are
attainable.

The possibility to use adiabatic passage for optical cooling was
demonstrated by Korsunsky~\cite{Korsunsky1996}, being applied to
\mbox{VSCPT}~\cite{Aspect1988} cooling. Raman cooling is a much
greater challenge which requires a careful analysis of the velocity
selection process. Another difficulty consists in a reasonable choice
of adiabatic passage which would demonstrate a benefit from its usage
in Raman cooling. For instance, Raman cooling employing rapid
adiabatic passage (RAP) was presented in
papers~\cite{Kuhn1996,Perrin1998,Perrin1999}, where the amplitude and
the frequency of Raman pulses are changed in a controlled way,
producing a velocity-selective transfer of the atomic population. The
extent of the frequency-chirp determines the range of excitation by
RAP and makes it possible to transfer even the wing of the velocity
distribution. However, RAP does not give an appreciable advantage over
ordinary Raman cooling, because the excitation profile depends on the
amount of chirping, the pulse area and its envelope. The theory
presented in this paper shows that the utilization of stimulated Raman
adiabatic passage (\mbox{STIRAP})~\cite{Gaubatz1990,Bergmann1998} in
contrast leads to a crucial growth of cooling robustness. In fact,
once the adiabaticity criterion is fulfilled, the velocity selection
in a case of large upper-level detuning depends only on the pulse
amplitudes, allowing wide variations in the pulse duration.

This paper is organized as follows. After describing the basic
atom-pulse setup in Sec.~\ref{sec:master}, we discuss the adiabaticity
criterion in Sec.~\ref{sec:resonance}. Assuming that this criterion is
satisfied, we present in Sec.~\ref{sec:ad_follow} the conditions of
the adiabatic conversion for internal atomic states required for
cooling. In Sec.~\ref{sec:v_select} we discuss the velocity selection
process and in Sec.~\ref{sec:cycle} the elementary cooling cycle. As
the full cooling process requires a series of these cycles, and the
adiabaticity criterion should be fulfilled for each cycle, we
formulate in Sec.~\ref{sec:cooling} two possible approaches: a) using
STIRAP pulses of equal amplitude, but with durations and amplitudes
that vary during the pulse cycle series (in analogy with the standard
Raman method, see e.g. Ref.~\cite{Ivanov2011}), and b) using STIRAP
pulses of substantially different amplitudes, but not changing their
duration and amplitude during the pulse cycle series. In the latter
case, the amplitudes are adjusted to transfer the wing of the velocity
distribution together with the possibility to cool the atomic ensemble
below the recoil limit. The cooling results for a series of cycles for
these two different transfer types are presented and discussed. The
paper is concluded by the summary given in Sec.~\ref{sec:final}.

\section{The atomic Hamiltonian}\label{sec:master}

\begin{figure}
\hbox to \linewidth {
\parbox[b]{4cm}{\includegraphics{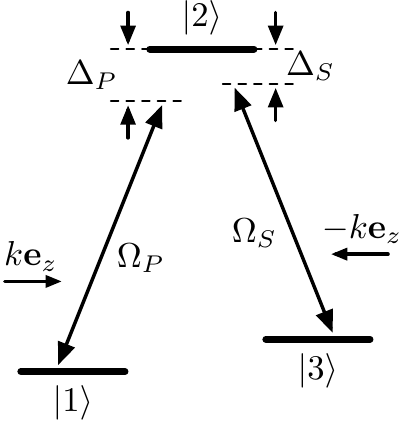}\\(a)}
\hfil
\parbox[b]{3.7cm}{\includegraphics{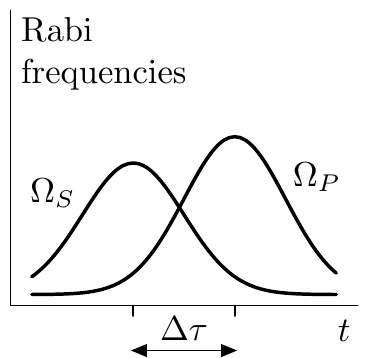}\\(b)}
}
\caption{\label{fig:lambda} (a) A three-level $\Lambda$-type atom
  interacting with a pair of contra-propagating laser beams. The pump
  laser links the initial state $|1\rangle$ with an intermediate state
  $|2\rangle$, whereas the Stokes laser links the intermediate state
  with the final state $|3\rangle$. (b) Both the pump and Stokes pulse
  have a Gaussian profile in time, overlapping during period
  $\Delta\tau$.}
\end{figure}

Figure \ref{fig:lambda}(a) illustrates a three-level $\Lambda$-type
atom that travels along direction $Oz$ with velocity $v$, being
originally prepared in ground state $|1\rangle$. A pump laser
propagating along the $Oz$ axis couples the atomic transition
$|1\rangle\leftrightarrow |2\rangle$, whereas the contra-propagating
Stokes laser couples the transition $|2\rangle\leftrightarrow
|3\rangle$. The electric field of laser pulses is written as
\begin{align*}
  \mathbf E(\mathbf{r}, t) = \frac{1}{2} \mathbf E_P e^{ikz -
    i\omega_P t} + \frac{1}{2} \mathbf E_S e^{-ikz -i\omega_S t} +
  \text{c.c.},
\end{align*} 
where $\omega_P$, $\omega_S$ are the corresponding laser frequencies,
and $k$ is the wave number for both light beams (the relevant
frequency quantities are in fact the atom-field detunings, which are
small compared to the actual frequencies, and thus we can assume equal
values of $k$). The pump and Stokes pulses produce a
velocity-selective \mbox{STIRAP} from the $|1\rangle$ to $|3\rangle$
ground state.

The total Hamiltonian for the atom-light system consists of the
kinetic-energy term $\hat{\mathbf P}^2/2M$, the Hamiltonian of the
$\Lambda$-type atom and the interaction Hamiltonian $\hat V$:
\begin{align*}
  \hat H = \frac{ \hat{\mathbf P}^2 }{2M} + \sum_j E_j |j\rangle
  \langle j| + \hat V.
\end{align*}
The laser-atom coupling $\hat V$ is the sum of couplings with both the
pump and Stokes pulse, and in the rotating-wave approximation (RWA) is
given by
\begin{multline*}
  \hat V_P = - \frac{d_{21} E_P}{2} |2\rangle \langle 1| e^{ikz
    -i\omega_P t}
  \\
  - \frac{d_{23} E_S}{2} |2\rangle \langle 3| e^{-ikz - i\omega_S t} +
  \mbox{H.c.},
\end{multline*} 
where $d_{21}$, $d_{23}$ are the corresponding electric dipole
moments; \mbox{H.c.} is the Hermitian conjugate. The Rabi
frequencies
\begin{align*}
  \Omega_P = - \frac{d_{21} E_P}{\hbar}, \quad \Omega_S = -
  \frac{d_{23} E_S}{\hbar}
\end{align*}
are considered to be real-valued and positive, and thus the
coupling $\hat V$ is written as
\begin{align}\label{eq:V}
  \hat V = \frac{\hbar\Omega_P}{2}|2\rangle \langle 1| e^{ikz
    -i\omega_P t} + \frac{\hbar\Omega_S}{2}|2\rangle \langle 3|
  e^{-ikz -i\omega_S t} + \mbox{H.c.}
\end{align}

Considering $z$ in Eq.~\eqref{eq:V} as an operator acting on the
external degrees of freedom of the atom, we apply in
Eq.~\eqref{eq:V} the relationship
\begin{align*}
  e^{\pm ikz} = \sum_p |p\rangle \langle p\mp \hbar k|.
\end{align*} 
The obtained laser-atom coupling
\begin{multline}\label{eq:V_sum}
  \hat V = \sum_p\left( \frac{\hbar\Omega_P}{2}|2,p\rangle\langle 1, p
    - \hbar k| e^{-i\omega_P t} \right.
  \\
  \left. \vphantom{ \frac{1}{2} } + \frac{\hbar\Omega_S}{2}|2,p\rangle
    \langle 3, p + \hbar k| e^{-i\omega_S t} \right) + \text{H.c.}
\end{multline}
shows that three states coupled by the pump and Stokes pulses
form a momentum family
\begin{align*}
  \mathcal F(p) = \{ |1,p - \hbar k\rangle, |2,p\rangle, |3,p + \hbar
  k\rangle \}.
\end{align*}
As long as spontaneous emission is not taken into account, a single
family $\mathcal F(p)$ of a specific momentum $p$ can be only taken
into account, so the sum in Eq.~\eqref{eq:V_sum} is avoided.

In the basis of the three states
\begin{align*}
  |a_1 \rangle &= \exp\left[-i\left(\frac{E_1}{\hbar} + \frac{(p-
        \hbar k)^2}{2M\hbar}\right) t\right] |1, p - \hbar k\rangle,
  \\
  |a_2 \rangle &= \exp\left[-i\left(\frac{E_2}{\hbar} + \frac{(p-
        \hbar k)^2}{2M\hbar} + \Delta_P \right) t\right] |2,p\rangle,
  \\
  |a_3 \rangle &= \exp\left[-i\left(\frac{E_3}{\hbar} + \frac{(p-
        \hbar k)^2}{2M\hbar} + \Delta_P - \Delta_S \right) t\right]
  |3, p + \hbar k\rangle,
\end{align*}
the dynamics of an atom starting from state $|a_1\rangle$ with the
corresponding velocity $v = (p - \hbar k)/M$ is described by the
atomic Hamiltonian
\begin{align}\label{eq:Ham}
  \tilde H = \hbar \begin{bmatrix} 0 & \dfrac{1}{2}\Omega_P(t) & 0
    \\
    \dfrac{1}{2}\Omega_P(t) & \dfrac{kp}{M} - \omega_R - \Delta_P &
    \dfrac{1}{2}\Omega_S(t)
    \\
    0 & \dfrac{1}{2}\Omega_S(t) & \dfrac{2kp}{M} + \Delta_S - \Delta_P
  \end{bmatrix},
\end{align}
where $\Delta_P = \omega_P - \omega_{21}$, $\Delta_S = \omega_S -
\omega_{23}$ are the pulse detunings; $\omega_R=\hbar k^2 /2M$ is the
recoil frequency.

The pump and Stokes pulses evolve in time as Gaussian profiles shown
in Fig.~\ref{fig:lambda}(b), being arranged in the counterintuitive
sequence with $t_S < t_P$:
\begin{align*}
  \Omega_P(t) = \Omega_{P0} e^{ -(t-t_P)^2/ 2T_P^2 }, \quad
  \Omega_S(t) = \Omega_{S0} e^{ -(t-t_S)^2/ 2T_S^2 },
\end{align*}
where $2T_P$, $2T_S$ are pulse widths. These lasers stimulate a
velocity-selective \mbox{STIRAP} from the $|a_1\rangle$ state to the
$|a_3\rangle$ state. The corresponding resonant velocity group is
considered in Sec.~\ref{sec:resonance}, whereas
Sec.~\ref{sec:ad_follow} and Sec.~\ref{sec:v_select} are devoted to an
arbitrary velocity group in the case of large enough detunings
$\Delta_P$ and $\Delta_S$.

\section{Two-photon resonance}\label{sec:resonance}

Two-photon resonance is characterized by momentum $p = M(\Delta_P -
\Delta_S)/2k$, for which the detuning from state $|3\rangle$ in
representation \eqref{eq:Ham} equals zero. Thus an atom prepared in
state $|1\rangle$ falls into the resonance, if it starts with resonant
velocity
\begin{align}\label{eq:v_res}
  v_0 = \frac{\Delta_P - \Delta_S}{2k} - v_\mathrm{rec},
\end{align}
where $v_\mathrm{rec}$ is the recoil velocity. Under this condition,
the basis of time-dependent eigenstates of the
Hamiltonian~\eqref{eq:Ham} is given by (see Ref.~\cite{Bergmann1998})
\begin{align}\label{eq:eigenst}
\begin{aligned}
  &|a^+\rangle = \sin\Theta \sin\Phi |a_1\rangle + \cos\Phi
  |a_2\rangle + \cos\Theta \sin\Phi |a_3\rangle,
  \\
  &|a^0\rangle = \cos\Theta |a_1\rangle - \sin\Theta |a_3\rangle,
  \\
  &|a^-\rangle = \sin\Theta \cos\Phi |a_1\rangle - \sin\Phi
  |a_2\rangle + \cos\Theta \cos\Phi |a_3\rangle.
\end{aligned}
\end{align} 
The corresponding (time-dependent) dressed-state eigenvalues are
\begin{align*}
\begin{aligned}
  \omega^+ &= \frac{1}{2} \left( \sqrt{ \tilde\Delta_P^2 +
      \Omega_P^2(t) + \Omega_S^2(t) } - \tilde\Delta_P \right), \quad
  \omega^0 = 0,
  \\
  \omega^- &= \frac{1}{2} \left( \sqrt{ \tilde\Delta_P^2 +
      \Omega_P^2(t) + \Omega_S^2(t) } + \tilde\Delta_P \right),
\end{aligned}
\end{align*}
where $\tilde\Delta_P = \Delta_P + \omega_R - kp/M$. The angle $\Phi$
is a function of the Rabi frequencies and detunings \cite{Fewell1997}:
\begin{align}\label{eq:phi}
  \tan\Phi = \frac{ \sqrt{\Omega_P^2(t) + \Omega_S^2(t)} } { \sqrt{
      \tilde\Delta_P^2 + \Omega_P^2(t) + \Omega_S^2(t) } -
    \tilde\Delta_P },
\end{align}
whereas the mixing angle $\Theta$ depends only on Rabi frequencies:
\begin{align}\label{eq:theta}
  \tan\Theta = \frac{\Omega_P(t)}{\Omega_S(t)}.
\end{align} 

In the case of large detuning $\Delta_P$ ($\Omega_P(t),\Omega_S(t) \ll
|\Delta_P|$), the basis of eigenstates $\{ |a^+\rangle, |a^0\rangle,
|a^-\rangle \}$ is reduced to the basis of states $\{ |\mathrm
C\rangle, |\mathrm{NC}\rangle, |a_2\rangle \}$. Both the coupled
$|\mathrm C\rangle$ and non-coupled $|\mathrm{NC}\rangle$ state is a
combination of ground states:
\begin{align}\label{eq:eigenst1}
\begin{aligned}
  &|\mathrm C\rangle = \sin\Theta |a_1\rangle + \cos\Theta
  |a_3\rangle,
  \\
  &|\mathrm{NC}\rangle = \cos\Theta |a_1\rangle - \sin\Theta
  |a_3\rangle,
\end{aligned}
\end{align} 
with the corresponding eigenvalues
\begin{align*}
  \omega_\mathrm{C} = \frac{ \Omega_P^2(t) + \Omega_S^2(t)
  }{4\Delta_P}, \quad \omega_\mathrm{NC} = 0.
\end{align*}
Here, we take into account that $|\Delta_P| \gg \omega_R, kp/M$.

The atomic population is contained in states $|\mathrm C\rangle$ and
$|\mathrm{NC}\rangle$, the Hamiltonian matrix element for nonadiabatic
coupling between these states is given by $\langle \mathrm C|
\frac{d}{dt}|\mathrm{NC}\rangle$ \cite{Messiah1962}. The ``local''
adiabaticity constraint reads that this matrix element should be small
compared to the field-induced energy splitting $|\omega_\mathrm{C} -
\omega_\mathrm{NC}|$, i.e.,
\begin{align*}
  | \langle \mathrm C| \frac{d}{dt}|\mathrm{NC}\rangle | \ll |
  \omega_\mathrm{C} - \omega_\mathrm{NC} |, \quad |\dot\Theta| \ll
  \frac{ \Omega_P^2(t) + \Omega_S^2(t) }{ 4|\Delta_P| }.
\end{align*}
Taking a time average of the left-hand side,
$\langle\dot\Theta_{av}\rangle=\pi/2\Delta\tau$, where $\Delta\tau$ is
the period during which the pulses overlap, one gets a convenient
``global'' adiabaticity criterion
\begin{align}\label{eq:ad_cr}
  \frac{ \Omega_P^2(t) + \Omega_S^2(t) }{|\Delta_P|}\Delta\tau
  \gg 1.
\end{align}

\section{The eigenstates of the effective Hamiltonian}\label{sec:ad_follow}

In the case of large detuning $\Delta_P$ ($\Omega_P(t),\Omega_S(t) \ll
|\Delta_P|$), adiabatic transfer can be considered at arbitrary
velocity. Then the upper state $|a_2\rangle$ is almost unpopulated
during \mbox{STIRAP} and can be adiabatically eliminated. Hence,
substituting the probability function $|\Psi\rangle$ into the
Schr\"odinger equation, one may assume that $\langle a_2 |
\frac{d}{dt} |\Psi \rangle \approx 0$. Contributions into state
$|a_2\rangle$ are derived from the Hamiltonian \eqref{eq:Ham} and, in
the case of $|\Delta_P| \gg \omega_R, kp/M$, are given by
\begin{align}\label{eq:ad_el}
  \langle a_2 |\Psi \rangle \approx \frac{\Omega_P(t)}{2\Delta_P}
  \langle a_1 |\Psi \rangle + \frac{\Omega_S(t)}{2\Delta_P} \langle
  a_3 |\Psi \rangle.
\end{align}
Approximation \eqref{eq:ad_el} requires that the condition
\begin{align}\label{eq:ad_el_cr}
  |\langle a_2 | \frac{d}{dt} |\Psi \rangle | \ll |\Delta_P| |\langle
  a_2 |\Psi \rangle|,
\end{align}
is fulfilled. By substituting Eq.~\eqref{eq:ad_el}
into the left-hand side of inequality \eqref{eq:ad_el_cr}, one gets
the necessary constraints
\begin{align*}
  |\Delta_S - \Delta_P|, T^{-1} \ll |\Delta_P|.
\end{align*}

Using Eq.~\eqref{eq:ad_el}, \mbox{STIRAP} can be described in the
basis of states $\{|a_1\rangle,|a_3\rangle \}$, and the effective
Hamiltonian of this two-level system is written as
\begin{align}\label{eq:Ham_eff}
  \hat H_\mathrm{eff}=\frac{\hbar}{2}\begin{pmatrix} -2\delta_0(t) &
    \Omega_\mathrm{eff}(t)
    \\
    \Omega_\mathrm{eff}(t) & 2(\delta_\mathrm{eff}(t) - \delta_0(t))
  \end{pmatrix}.
\end{align} 
The effective detunings and the Rabi frequency are
\begin{align}\label{eq:var_eff}
  \begin{aligned}
    & \delta_0(t) = -\frac{\Omega_P^2(t)}{4\Delta_P}, \quad
    \Omega_\mathrm{eff}(t) = \frac{\Omega_P(t)
      \Omega_S(t)}{2\Delta_P},
    \\
    & \delta_\mathrm{eff}(t) = \Delta\delta + \frac{\Omega_S^2(t) -
      \Omega_P^2(t)}{4\Delta_P},
  \end{aligned}
\end{align}
where $\Delta\delta = \Delta_S - \Delta_P + 2kp/M = 2k (v - v_0)$; $v$
is the original velocity of an atom along axis $Oz$. The effective
Hamiltonian~\eqref{eq:Ham_eff} has the following eigenstates:
\begin{align}\label{eq:eigen_st}
\begin{aligned}
  |a_+\rangle &= \sin\Theta |a_1\rangle + \cos\Theta |a_3\rangle,
  \\
  |a_-\rangle &= \cos\Theta |a_1\rangle - \sin\Theta |a_3\rangle,
\end{aligned}
\quad \tan\Theta = \sqrt{ 1 +
  \frac{\delta_\mathrm{eff}^2}{\Omega_\mathrm{eff}^2} } -
\frac{\delta_\mathrm{eff}}{\Omega_\mathrm{eff}},
\end{align} 
with the corresponding eigenfrequencies
\begin{align*}
  \omega_\pm = -\delta_0 + \frac{\delta_\mathrm{eff}}{2} \pm
  \frac{\Omega_\mathrm{eff}}{2} \sqrt{ 1 +
    \delta_\mathrm{eff}^2/\Omega_\mathrm{eff}^2 }.
\end{align*}

\begin{figure}
\includegraphics[width=6.5cm]{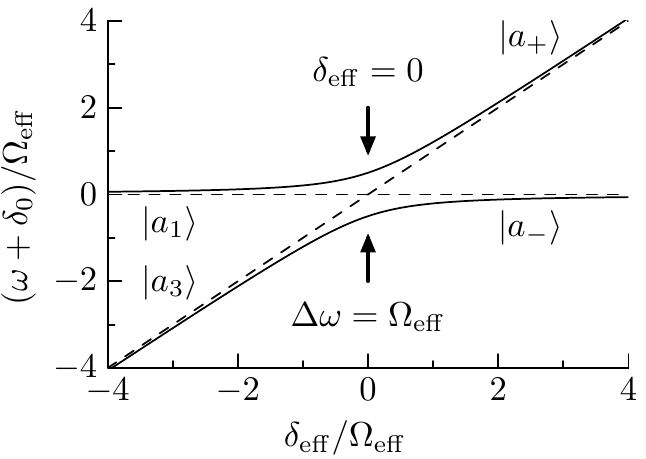}
\caption{\label{fig:dressed_st} The eigenfrequencies of pure atomic
  states $|a_1\rangle$ and $|a_3\rangle$ (dashed lines) cross at point
  $\delta_\mathrm{eff}=0$. The avoided level crossing for the
  eigenstates (solid lines) leads to a corresponding change between
  the eigenstates and pure states.}
\end{figure}

The approach of eigenstates $|a_+\rangle$ and $|a_-\rangle$ covers the
case of two-photon resonance \eqref{eq:v_res} as a specific case with
$\Delta\delta = 0$. In this case, the mixing angle $\Theta$ in
Eq.~\eqref{eq:eigen_st} coincides with that in Eq.~\eqref{eq:theta},
and eigenstates $|a_+\rangle$, $|a_-\rangle$ turn into states
$|\mathrm C\rangle$ and $|\mathrm{NC}\rangle$, respectively. During
\mbox{STIRAP}, the angle $\Theta$ varies from $0$ to $\pi/2$, and the
non-coupled state $|\mathrm{NC}\rangle$ evolves from the $|a_1\rangle$
to $|a_3\rangle$ state, involving an adiabatic transfer of
population. Consequently, the eigenvalue $\omega_\mathrm{NC}$ changes
from the frequency of state $|a_1\rangle$ to that of state
$|a_3\rangle$.

General case of arbitrary $\Delta\delta$ can be considered in the same
way where the role of state $|\mathrm{NC}\rangle$ is played by the
eigenstate $|a_-\rangle$. \mbox{STIRAP} occurs together with a change
in the eigenfrequencies $\omega_\pm$, whose dependence on detuning
$\delta_\mathrm{eff}$ is shown in Fig.~\ref{fig:dressed_st}. While the
laser pulses overlap, $\delta_\mathrm{eff}$ can vary from negative to
positive value or vice versa, in dependence on the sign of
$\Delta_P$. The switching point between states $|a_1\rangle$ and
$|a_3\rangle$ corresponds to a crossing between the frequencies of
these pure states, which appears at $\delta_\mathrm{eff} = 0$.

\begin{figure}
\centering
\includegraphics[width=6.5cm]{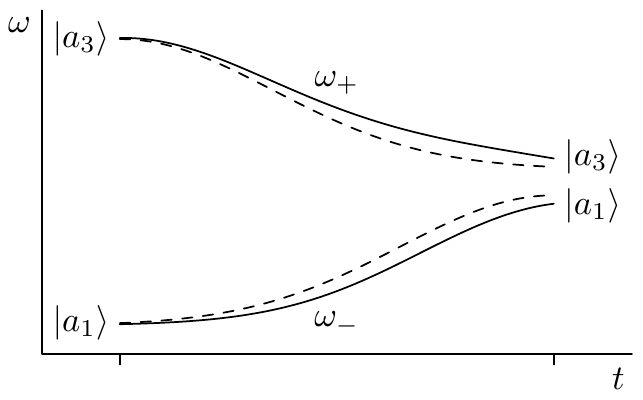}(a)\\
\includegraphics[width=6.5cm]{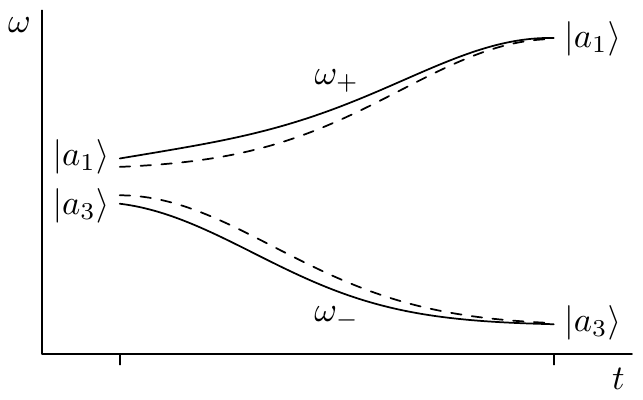}(b)\\
\includegraphics[width=6.5cm]{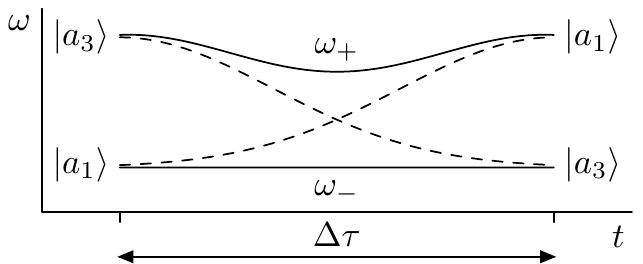}(c)
\caption{\label{fig:var_delta}
  The time evolution of the dressed-state frequencies (solid line) on
  the pulse-overlapping region $\Delta\tau$. The dashed lines
  represent the frequencies of the pure states. In dependence on
  $\Delta\delta$, one gets the following cases: (a)
  $\delta_\mathrm{eff} > 0$, (b) $\delta_\mathrm{eff} < 0$ during this
  period; (c) $\delta_\mathrm{eff} = 0$ at a specific time. (c)
  represents adiabatic transfer from the $|a_1\rangle$ state to
  $|a_3\rangle$ state.  }
\end{figure}

The time evolution of the eigenfrequencies $\omega_\pm$ is shown in
Fig.~\ref{fig:var_delta} where three possible cases against the
detuning $\delta_\mathrm{eff}$ are presented: (a) $\delta_\mathrm{eff}
> 0$ or (b) $\delta_\mathrm{eff} < 0$ all the time while laser pulses
overlap; (c) $\delta_\mathrm{eff} = 0$ at a specific time during this
period. Both eigenstates $|a_+\rangle$ and $|a_-\rangle$ in
Figs.~\ref{fig:var_delta}(a) and \ref{fig:var_delta}(b) return to the
initial pure state in the end of \mbox{STIRAP}, suppressing adiabatic
transfer of atoms. A crossing between the frequencies of states
$|a_1\rangle$ and $|a_3\rangle$ appears in
Fig.~\ref{fig:var_delta}(c), where atoms that adiabatically follow the
eigenstate $|a_-\rangle$ are transferred from the $|a_1\rangle$ to
$|a_3\rangle$ state. The corresponding velocity group derived from the
explicit form \eqref{eq:var_eff} of $\delta_\mathrm{eff}$ is given by
\begin{align}\label{eq:v_range}
\begin{aligned}
  -\frac{\Omega_{S0}^2}{8k|\Delta_P|} < v - v_0 <
  \frac{\Omega_{P0}^2}{8k|\Delta_P|} \quad \mbox{if $\Delta_P > 0$},
  \\
  -\frac{\Omega_{P0}^2}{8k|\Delta_P|} < v - v_0 <
  \frac{\Omega_{S0}^2}{8k|\Delta_P|} \quad \mbox{if $\Delta_P < 0$}.
\end{aligned}
\end{align}

If Rabi frequency $\Omega_\mathrm{eff}(t)$, detunings
$\delta_\mathrm{eff}(t)$ and $\delta_0(t)$ vary sufficiently slowly so
that the nonadiabatic mixing between states $|a_+\rangle$ and
$|a_-\rangle$ is negligible, population remains in state $|a_-\rangle$
during the whole \mbox{STIRAP} pulse. If this condition holds for all
velocity groups, the velocity range \eqref{eq:v_range} corresponds to
population transfer from state $|a_1\rangle$ to $|a_3\rangle$. The
transfer is accompanied by the probability near unity for all atoms
within the range \eqref{eq:v_range}, is not affecting outside
atoms. However, because the transfer should be adiabatic for huge
velocities, required pulse duration is so long that an acceptably
efficient cooling can not be achieved. In contrast, we suggest the
adiabaticity criterion to be satisfied only for atoms at the
two-photon resonance \eqref{eq:v_res}, which in turn allows shorter
durations of the \mbox{STIRAP} pulse. The corresponding resonant
velocity range for \mbox{STIRAP} is considered next in the next
section.

\section{Velocity selection of \mbox{STIRAP} pulse}\label{sec:v_select}

In the case of large detuning $\Delta_P$, the state vector
$|\Psi\rangle$ follows the dressed state $|a_-\rangle$ adiabatically
throughout the interaction if a condition
\begin{align*}
  | \langle a_+| \frac{d}{dt} |a_-\rangle | \ll |\omega_- - \omega_+|,
\end{align*}
is fulfilled during the \mbox{STIRAP} pulse. After the substitution of
states $|a_+\rangle$, $|a_-\rangle$ and the corresponding
eigenfrequencies from Eq.~\eqref{eq:eigen_st}, this constraint is
written as
\begin{align}\label{eq:ad_cr_eff}
  |\dot\Theta| \ll \sqrt{\Omega_\mathrm{eff}^2+\delta_\mathrm{eff}^2}.
\end{align}
Next we consider the ``global'' adiabaticity criterion, replacing
$\dot\Theta$ in Eq.~\eqref{eq:ad_cr_eff} by an appropriate time average
$\langle\dot\Theta_{av}\rangle$.

The rate of population transfer from state $|a_1\rangle$ to
$|a_3\rangle$ is the largest for atoms in the two-photon resonance,
which is valid regardless of whether the transfer is adiabatic or
not. If atoms follow the eigenstate $|a_-\rangle$ adiabatically, both
population in state $|a_3\rangle$ and the mixing angle $\Theta$ has
the similar behaviour, because the population varies as
$\sin^2\Theta$. We assume that $\Theta$ has the largest rate for
resonant atoms as well, at least for quite large
$\Delta\tau$. This fact relative to an average value
$\langle\dot\Theta_{av}\rangle$ gives an inequality
\begin{align*}
  \langle \dot\Theta_{av} \rangle \le \pi/2 \Delta\tau.
\end{align*}
In cases shown in Figs.~\ref{fig:var_delta}(a) and
\ref{fig:var_delta}(b), in spite of the fact that each atomic
population at the beginning and the end of laser pulse takes the same
value, $\Theta$ can change significantly in range from $0$ to
$\pi/4$. Hence $\langle\dot\Theta_{av}\rangle$ can be of the same
order of magnitude as $\Delta\tau^{-1}$.

After the substitution of $\delta_\mathrm{eff}(t)$,
$\Omega_\mathrm{eff}(t)$, to the inequality \eqref{eq:ad_cr_eff}, the
adiabaticity criterion reads
\begin{align}\label{eq:ad_cr_eff1}
  \frac{1}{\Delta\tau} \ll \sqrt{ \left( \frac{\Omega_P^2(t) +
        \Omega_S^2(t)}{4\Delta_P} \right)^2 + \Delta\delta
    \left(\Delta\delta + \frac{\Omega_S^2(t) -
        \Omega_P^2(t)}{2\Delta_P}\right) }.
\end{align}

The two-photon resonance corresponds to $\Delta\delta = 0$, when the
inequality \eqref{eq:ad_cr_eff1} coincides with the adiabaticity
criterion \eqref{eq:ad_cr}. As seen from Eq.~\eqref{eq:ad_cr_eff1},
once the adiabaticity criterion \eqref{eq:ad_cr} is fulfilled, atoms
following state $|a_-\rangle$ adiabatically are given by
\begin{align*}
  \Delta\delta \left(\Delta\delta + \frac{\Omega_S^2(t) -
      \Omega_P^2(t)}{2\Delta_P}\right) \ge 0.
\end{align*}
Hence, in addition to the resonance-velocity group, adiabatic transfer
includes the following velocity groups:
\begin{align*}
  \begin{aligned}
    v - v_0 \le -\frac{\Omega_{S0}^2}{4k|\Delta_P|} \quad \mbox{or}
    \quad v - v_0 \ge \frac{\Omega_{P0}^2}{4k|\Delta_P|} \quad
    \mbox{if $\Delta_P > 0$},
    \\
    v - v_0 \le -\frac{\Omega_{P0}^2}{4k|\Delta_P|} \quad \mbox{or}
    \quad v - v_0 \ge \frac{\Omega_{S0}^2}{4k|\Delta_P|} \quad
    \mbox{if $\Delta_P < 0$}.
  \end{aligned}
\end{align*}
However, in this case, $\Theta$ equals $0$ at the end of \mbox{STIRAP}
pulse, and atoms return to state $|a_1\rangle$. That is why only those
atoms can be transferred to state $|a_3\rangle$, for which
\begin{align}\label{eq:v_range1}
  \begin{aligned}
    -\frac{\Omega_{S0}^2}{4k|\Delta_P|} < v - v_0 <
    \frac{\Omega_{P0}^2}{4k|\Delta_P|} \quad \mbox{if $\Delta_P > 0$},
    \\
    -\frac{\Omega_{P0}^2}{4k|\Delta_P|} < v - v_0 <
    \frac{\Omega_{S0}^2}{4k|\Delta_P|} \quad \mbox{if $\Delta_P < 0$}.
  \end{aligned}
\end{align}
The velocity range \eqref{eq:v_range1} is only defined by the
two-photon Rabi frequencies, being twice broader than that by
Eq.~\eqref{eq:v_range}. The transfer probability reaches unity at the
two-photon resonance and then declines to zero as the boundary of range
\eqref{eq:v_range1} approaches. Possible velocity envelopes will be
discussed in Sec.~\ref{sec:cooling} as applied for Raman cooling.

The \text{STIRAP} pulse transports population from the $|1\rangle$ to
$|3\rangle$ state without an occupation of the immediate state
$|2\rangle$ for the resonant-group atoms only. For the rest of atoms,
the transfer is accompanied by populating the $|2\rangle$ state, which
in turn gives rise to a spontaneous decay from this state to ground
states of the $\Lambda$-type atom.

A contribution from spontaneous decay can be estimated with help of
the density operator $\sigma$, whose matrix elements are given by
\begin{align}\label{eq:sigma}
  \sigma_{ij}(p) = \langle a_i |\sigma| a_j \rangle, \quad i,j =
  1,2,3.
\end{align}
Portions of atoms in state $|a_2\rangle$ and those of them leaving
state $|a_2\rangle$ due to spontaneous decay, $\sigma_{22}(p)$ and
$\sigma_\mathrm{sp}(p)$, are related by relationship
\begin{align*}
  \frac{d}{dt}\sigma_\mathrm{sp}(p) = \Gamma \sigma_{22}(p),
\end{align*}
where $\Gamma$ is the rate of spontaneous emission. Because state
$|a_2\rangle$ is only populated during period $\Delta\tau$, while the
pump and Stokes pulses overlap, $\sigma_\mathrm{sp}(p)$ is given by
\begin{align}\label{eq:sp_decay}
  \sigma_\mathrm{sp}(p) = \Gamma \int_0^{\Delta\tau} \sigma_{22}(p) \,
  dt.
\end{align}
The inequality \eqref{eq:ad_el} imposes a constraint on a fraction of
atoms in state $|a_2\rangle$:
\begin{align*}
  \sigma_{22}(p) \lesssim \frac{\Omega_P^2(t)}{2\Delta_P^2}
  \sigma_{11}(p) + \frac{\Omega_S^2(t)}{2\Delta_P^2} \sigma_{33}(p)
  \le \frac{\Omega_P^2(t) + \Omega_S^2(t)}{2\Delta_P^2}.
\end{align*}
Substituting $\sigma_{22}(p)$ into Eq.~\eqref{eq:sp_decay}, one gets a
contribution of the spontaneous decay
\begin{align*}
  \sigma_\mathrm{sp}(p) \lesssim \Gamma \frac{\Omega_P^2(t) +
    \Omega_S^2(t)}{2\Delta_P^2} \Delta\tau.
\end{align*}
Atoms spontaneously decaying from state $|2\rangle$ are neglected from
the consideration if $\sigma_\mathrm{sp}(p) \ll 1$. This condition in
combination with the adiabaticity criterion \eqref{eq:ad_cr} gives
the constraint
\begin{align*}
  \frac{|\Delta_P|}{\Gamma} \gg \frac{\Omega_P^2(t) +
    \Omega_S^2(t)}{|\Delta_P|} \Delta\tau \gg 1,
\end{align*}
which can be reached by increasing detuning $\Delta_P$.

\section{Elementary cooling cycle}\label{sec:cycle}

\begin{figure}
\includegraphics[width=6.9cm]{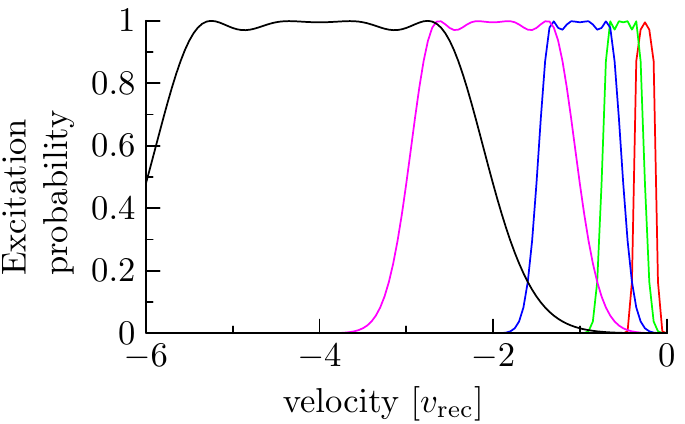}(a)\\
\includegraphics[width=6.75cm]{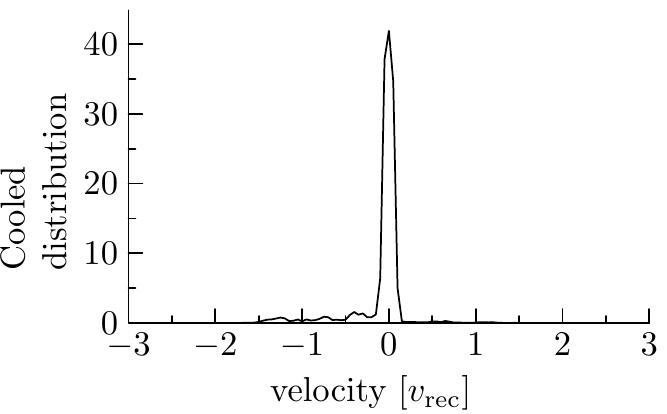}(b)
\caption{\label{fig:set}
  (Color online) (a) The excitation probability of atoms in state
  $|1\rangle$ against velocity $v$ by \mbox{STIRAP} pulses of
  different coupling parameters according to set \eqref{eq:v_set}. (b)
  The velocity spread $\Delta v$ decreases from the initial
  $3v_\mathrm{rec}$ to $0.1v_\mathrm{rec}$ of cooled ensemble.  }
\end{figure}

The first step of a cooling cycle consists in transferring atoms from
state $|1\rangle$ to $|3\rangle$ due to \text{STIRAP}
pulse. Once the adiabaticity criterion \eqref{eq:ad_cr} is fulfilled,
atoms at the resonant velocity $v_0$ \eqref{eq:v_res} are transferred
with efficiency near unity. For laser configuration depicted in
Fig.~\ref{fig:lambda}, atoms after the transfer get a momentum gain of
$2\hbar k$ along $Oz$, so one should follow with a condition
$v_0<0$. Because the spread of the resonant-velocity group is defined
by Eq.~\eqref{eq:v_range1}, atoms at zero velocity are suppressed from
the transfer when
\begin{align}\label{eq:v0_range}
  v_0 \le -\dfrac{\Omega_{P0}^2}{4k|\Delta_P|} \quad \mbox{if
    $\Delta_P > 0$}, \quad v_0 \le
  -\dfrac{\Omega_{S0}^2}{4k|\Delta_P|} \quad \mbox{if $\Delta_P < 0$}.
\end{align}
The inequality \eqref{eq:v0_range} only imposes a constraint on
maximal magnitude of Rabi frequencies and thereby of the velocity
spread of the resonant atoms. By decreasing Rabi frequency
$\Omega_{S0}$ or $\Omega_{P0}$, the magnitude of the resonant velocity
can be chosen substantially smaller than the recoil velocity. As a
result, \mbox{STIRAP} can be efficiently used as the first step of an
elementary cooling cycle. But such a transfer does not need the exact
holding of the pulse duration, as that does in the standard Raman
cooling \cite{Kasevich1992}.

In the second step, optical pumping excites atoms from $|3\rangle$ to
$|2\rangle$ state, changing the momentum of an atom from $p'$ to
$p'-\hbar k$. Then the atom spontaneously decays in the initial state
$|1\rangle$, and its momentum becomes $p=p'-\hbar k-\Delta p$ where
$\Delta p$ is the projection of a spontaneously emitted photon;
$|\Delta p| \le \hbar k$. The resulting population in state
$|1\rangle$ is written as
\begin{align*}
  \langle 1,p| \sigma' |1,p\rangle = \langle 1,p| \sigma |1,p\rangle +
  \langle 3,p {+} \hbar k {+} \Delta p| \sigma |3,p {+} \hbar k {+}
  \Delta p\rangle.
\end{align*}
In turn, the density-matrix element $\sigma_{ij}(p)$~\eqref{eq:sigma}
corresponds to the momentum family $\mathcal F(p)$ that includes
states $|a_i\rangle$ and $|a_j\rangle$, which leads to expression
\begin{align}\label{eq:sp_decay}
  \sigma'_{11}(p) = \sigma_{11}(p) + \sigma_{33}(p - \hbar k + \Delta
  p),
\end{align}
where the momentum shift is given against $\mathcal F(p)$.
Equation~\eqref{eq:sp_decay} demonstrates the mixing of different
families $\mathcal F(p)$ due to spontaneous decay. After the cooling
cycle the number of atoms at zero velocity increases together with the
cooling of the atomic ensemble.

\section{Comparison of two cooling types}\label{sec:cooling}

\begin{figure}
\includegraphics[width=6.9cm]{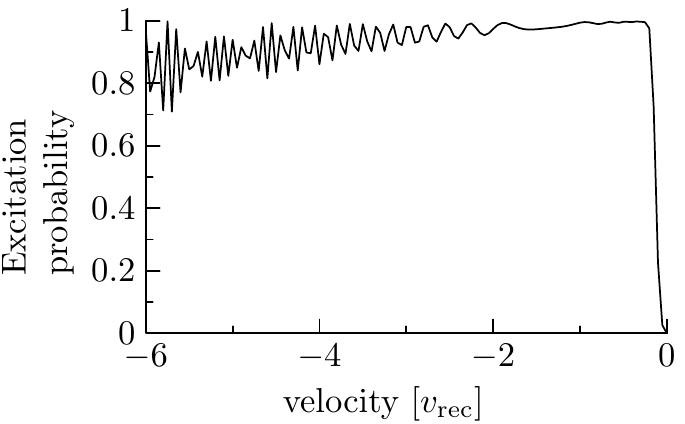}(a)\\
\includegraphics[width=6.75cm]{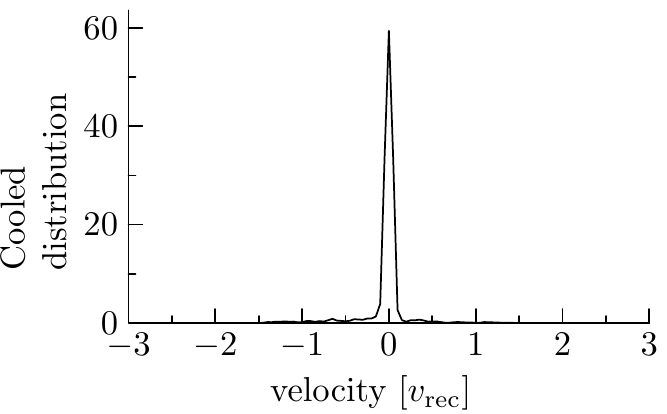}(b)
\caption{\label{fig:total}
  (a) The excitation probability of atoms in state $|1\rangle$ against
  velocity $v$ by \mbox{STIRAP} pulse \eqref{eq:asym_neg} with the
  maximum at the resonant velocity $v_0 = -0.25v_\mathrm{rec}$. (b)
  The velocity spread $\Delta v$ decreases from the initial
  $3v_\mathrm{rec}$ to $0.1v_\mathrm{rec}$ of cooled ensemble.  }
\end{figure}

In this section, we compare two different variants of cooling by
\text{STIRAP}. The first one uses the similar laser beams with
$\Omega_{P0} = \Omega_{S0}$, so the velocity profile of the pulse has
a symmetric shape. A condition when atoms at zero velocity are not
transferred is derived from Eq.~\eqref{eq:v0_range}, and is written as
\begin{align*}
  \Omega_{P0}^2 = \Omega_{S0}^2 = 4 k |v_0| |\Delta_P|.
\end{align*}
Both the pump and the Stokes pulse has the same pulse shape, so
\begin{align*}
T_P = T_S = 0.5 (t_P - t_S).
\end{align*}
As the resonant velocity $v_0$ gets closer to zero, the width of the
excitation profile decreases. To entirely excite the wing of the
initial profile of spread $3 v_\text{rec}$, we use a set of $v_0$
\begin{align}\label{eq:v_set}
  |v_0| = 4 v_\text{rec} / 2^k, \quad k = 0, \dots, 4,
\end{align}
for which the probability of transfer from state $|1\rangle$ is shown
in Fig.~\ref{fig:set}(a). Note that as $|v_0|$ decreases, the magnitudes
$\Omega_{P0}^2/|\Delta_P|$ and $\Omega_{S0}^2/|\Delta_P|$ decrease as
well, which in turn leads to an increase of $\Delta\tau$ in order to
fulfil the adiabaticity criterion \eqref{eq:ad_cr}. After the set has
been applied, laser beams are alternated in order to excite the
right-side profile. Such a cooling method is similar to ordinary Raman
cooling.

In the second variant, the complete wing of initial distribution
except a narrow peak at zero velocity is transferred by one
\text{STIRAP} pulse as shown in Fig.~\ref{fig:total}(a). The peak
width is defined by the resonant velocity $v_0$ which is equal to
$-0.25 v_\mathrm{rec}$. Notice that such an excitation is not
attainable for ordinary Raman cooling, and represents a new type of a
velocity-selective transfer with an essentially asymmetric
profile. The advantage is in utilizing a single profile instead of the
set \eqref{eq:v_set}. Rabi frequencies are given by
\begin{subequations}
\begin{align}
  \Omega_{P0}^2 = 4 k |v_0| |\Delta_P|, \Omega_{S0} = 8 \Omega_{P0}
  \quad \mbox{if $\Delta_P > 0$},\label{eq:asym_neg}
  \\
  \Omega_{S0}^2 = 4 k |v_0| |\Delta_P|, \Omega_{P0} = 8 \Omega_{S0}
  \quad \mbox{if $\Delta_P < 0$}.
\end{align}
\end{subequations}
Because a pulse of the largest amplitude should decrease faster,
pulse widths are given by
\begin{align*}
  T_P = 0.5 (t_P - t_S), T_S = 0.35 (t_P - t_S) \quad \mbox{if
    $\Delta_P > 0$},
  \\
  T_P = 0.35 (t_P - t_S), T_S = 0.5 (t_P - t_S) \quad \mbox{if
    $\Delta_P < 0$}.
\end{align*}
After each \mbox{STIRAP} pulse, laser beams are alternated giving the
same excitation of the right side of the velocity distribution.

In both cooling variants, the duration $T_\mathrm{pulse}$ of
\text{STIRAP} pulse is defined by the start and end time
\begin{align*}
  t_\mathrm{start} = t_S - (t_P - t_S), \quad t_\mathrm{end} = t_P +
  (t_P - t_S),
\end{align*}
being equal to $3(t_P - t_S)$. Pulse durations in the first variant in
accordance with the set \eqref{eq:v_set} are given by
\begin{align*}
  T_\mathrm{pulse} = 6 \cdot 2^k \tau_R \quad k = 0, \dots, 4,
\end{align*}
where $\tau_R = \omega_R^{-1}$ is the recoil time. In the second
variant, each pulse has the same duration equal to $96 \tau_R$.

Figures \ref{fig:set}(b) and \ref{fig:total}(b) show results of both
cooling methods starting from the initial distribution with the spread
of $3v_\mathrm{rec}$. The velocity spread $\Delta v = (\mathrm{FWHM})
/ \sqrt{8 \ln 2}$ has been reduced to nearly a) $0.06 v_\mathrm{rec}$
in Fig.~\ref{fig:set}(b), and b) $0.04 v_\mathrm{rec}$ in
Fig.~\ref{fig:total}(b). The corresponding temperatures of the atomic
ensemble have gone down to a) $0.004 T_\mathrm{rec}$, and a) $0.002
T_\mathrm{rec}$, where $T_\mathrm{rec}$ is the recoil-limit
temperature. The duration of optical pumping is considered as a
negligible value against that of \text{STIRAP} pulse, which in turn
gives total durations of cooling methods as $7440\tau_R$ and
$7680\tau_R$, respectively. In spite of the fact that both durations
are approximately equal, the number of cooling cycles are rather
different: the first method contains $N = 200$ elementary cycles,
whereas the second one contains $N = 80$ cycles. For instance, being
applied for cesium atoms ($\omega_R \sim 2\pi \times 2 \mbox{kHz}$),
both cooling variants take a time of $0.6 \mathrm s$.

\section{Conclusion}\label{sec:final}

We have described Raman cooling by velocity-selective \mbox{STIRAP} in
the case of large upper-level detuning. We assume that the
adiabaticity criterion is fulfilled for atoms in the two-photon
resonance, so these atoms are transferred with efficiency that
approaches unity. The position and width of the excitation profile is
insensitive to the pulse duration, being in a linear dependence on the
two-photon Rabi frequencies. This profile can take an asymmetric form,
exciting the wing of the velocity distribution with the exception of a
narrow peak near zero-velocity group. We have compared such a type of
\mbox{STIRAP} with that of symmetric profile where a set of different
coupling parameter utilized for excitation of the wing of the velocity
distribution. Both methods has well-controlled the excitation
probability, making the attainable temperatures essentially go below
the one-photon recoil limit.

Our simplified treatment does not necessarily take into account many
aspects of an actual experiment, but it shows that {\it a priori} the
idea of Raman cooling combined with \mbox{STIRAP} offers a tool for
reaching ultracold temperatures without the use of evaporative
cooling. This is an example of robust quantum control of translational
atomic degrees of freedom. If subjected to trapped atoms, 1D cooling
eventually leads to cooling at higher dimensions, although at tight
confinement the quantization of motion opens the door for sideband
cooling (which also relies on a Raman
process)~\cite{Ye2008,Kerman1999}. It can also be applied to
low-dimensional systems prepared with optical
lattices~\cite{Bloch2005}, such as the elongated quasi-1D cigar-shaped
clouds in which the lattice provides tight confinement in
perpendicular direction but atoms are almost free to move in axial
direction, in which one can then apply Raman cooling. If using such a
system as a wave guide, the trapping potential in axial direction is
quite weak, which limits the use of side-band cooling in axial
direction. Also, for atomic clocks one needs high numbers of atoms and
yet low densities to avoid interaction-induced frequency shifts, and
one solution is again an optical lattice that slices the sample into
non-interacting 2D pancakes~\cite{Derevianko2011}. Although then
narrow-band cooling is the basic tool for e.g. alkaline earth atoms,
one can also consider cooling at higher-lying atomic energy levels,
which gives a rich level structure and opens the possibility for Raman
cooling with tripod configuration. Finally, we note that the very
recent interest in sub-Doppler cooling of fermionic
isotopes~\cite{Gokhroo2011,Landini2011} provides yet another
application for non-evaporative cooling methods.

\section{Acknowledgments}

This research was supported by the Finnish Academy of Science and
Letters, CIMO, the Academy of Finland, grant 133682, by the Ministry of
Education and Science of the Russian Federation, grant RNP 2.1.1/2166,
and by the Russian Foundation for Basic Research, grant RFBR
09-02-00223-a.

\end{document}